\documentstyle[12pt,prb,epsfig,aps]{revtex}
\renewcommand{\baselinestretch}{1.0}
\begin{document}
\title{	Mechanisms of spatial current-density instabilities in
	$p^+$-$p^-$-$n$-$p^+$-$n^{++}$ structures}
\author{A.V. Gorbatyuk}
\address{A.F. Ioffe Physicotechnical Institute,
	Russian Academy of Science,
	Politechnicheskaya 26, 194021 St. Petersburg, Russia}
\author{F.-J. Niedernostheide}
\address{Institut f\"ur Angewandte Physik, Universit\"at M\"unster, 
	Corrensstra\ss e 2/4, 48149 M\"unster, Germany, and}
\address{Siemens AG, 81730 M\"unchen, Germany}
\date{accepted for publication in Phys. Rev. B}
\maketitle

\begin{abstract}
Semiconductor $p^+$-$p^-$-$n$-$p^+$-$n^{++}$ 
structures with  large junction and contact areas 
are treated as 1$\times$2-dimensional active media,
in which self-organized pattern formation can be expected.
The local bistable behavior of the structures may emanate
from two different mechanisms both  
governed by a nonlinear current feedback-loop 
between the electrons and holes
injected from the outer layers. 
By considering the device to be composed of
an active subsystem with negative differential resistance and 
a passive resistive layer with positive differential resistance
an analytical approach is suggested to understand and describe the 
corresponding physical mechanisms in a self-consistent way.  
Analytical solutions of  
the derived model equations allow a description of 
homogeneous stationary states and yield
explicit expressions of the 
current-density vs. voltage characteristics of the whole structure and 
its subsystems.
A stability analysis of the homogeneous states with
respect to two-dimensional transversal harmonic
fluctuations is performed for the two cases under study. 
The resulting dispersion relations allow
two different types of instability. 
While the first one is of Ridley's type which is characteristic for any 
spatially extended electrical system with negative differential 
resistance, the second type can be considered as a solid-state analogue 
of Turing's instability known as a generic instability mechanism
which may lead, e.\,g., to the formation of periodic patterns.
\end{abstract}

\section{Introduction}
Basic ideas on instabilities and dissipative pattern formation 
in open spatially extended nonlinear systems \cite{Tur,Prig,Hak}
have gained substantial interest in many fields of science 
in the last two decades.
(see, e.\,g., Refs.~\cite{Keros,Mich,Wal,Eng,Buss}).
The investigations become rather intensive also in the 
field of solid-state physics and electronics \cite{Keros,Sch87,Nied} 
where a large variety of different nonlinear 
nonequilibrium phenomena occurs, 
which can cause  spontaneously arising spatio-temporal patterns 
in materials and devices. 
Together with highly developed 
device technology, these circumstances suggest promising 
opportunities for future research and developements.   

Recently, attention has been called to thyristor-like 
semiconductor structures with large active areas
\cite{Nie92,Gor1,Gor2,GN1}, as 
these nonlinear systems with bistable properties 
show several spatial and spatio-temporal 
current-density patterns.   
Such semiconductor structures could potentially be used
as multi-stable elements for integrated circuits,
self-organizing devices for image recognition and
image processing, etc.~\cite{Hak,Bode,Ruw}.
One of the most striking results of former
considerations based on phenomenological models \cite{Nie92,Gor2}  
is that the instability mechanisms   
posess some features very similar to those studied in
biology or chemical media, which may show an extremly
manifold selforganizing behavior.      
However, at present the understanding of instability 
mechanisms and pattern formation in multilayer structures 
is far from being complete. 
Apparently, there is  a lack of systematic
investigations concerning self-organization aspects for
this family of devices. 
In a very early theoretical paper \cite{Varl} 
the process of single current filament formation has only been analyzed 
for the simplest case of $p$-$n$-$p$-$n$ structures, 
that according to Ridley's terminology \cite{Rid}                          
belong to the class of extended electrical systems 
with current controlled negative differential resistance.
Recently, gate-driven $p$-$n$-$p$-$n$ structures have been
used to study current filamentation and the 
dynamics of switching fronts under the constraints
of two global couplings \cite{Zphys,Meix98}. In specially designed
multilayer structures different types of filament oscillation
have been observed \cite{Nie92b,Nie94b,Nie97}.
Very few theoretical papers have been devoted to instabilities 
in standard thyristors; 
they all aimed merely on a safe operation in power applications
\cite{Prepr,Wach,Jae}. 
Furthermore, it is not an easy task to adapt 
the advanced experience of power device technology directly to 
the specifics of devices showing phenomena of self-organization.
One of the main reasons for this is that the operation
of semiconductor devices and especially thyristors depends 
very sensitively on many physical and design parameters.
On the one hand, this gives numerous opportunities
for a proper choice of those parameters
responsible for the intended function of these devices.
But on the other hand 
the high-dimensional parameter space
makes a systematic analysis rather difficult, because
the use of numerical methods and the performance of
experiments are usually time-consuming and/or expensive. 
Therefore detailed studies of the underlying physical processes 
in conjunction with the material and the design parameters 
are necessary and should lead to
the deepest possible analytical description of the device physics.

Following this motivation and aiming at further development
of concepts introduced earlier in Refs.~\cite{Nie92,Wier,GN1}
we suggest in this paper an analytical approach for instability mechanisms 
in a family of devices based on thyristor-like  
$p^+$-$p^-$-$n$-$p^+$-$n^{++}$ structures, 
in which several modes of pattern formation have already been
found experimentally \cite{Nie92}. 

The outline of the paper is as follows: In section 2 the investigated
semiconductor structure and its basic physics with emphasis on the
nonlinear properties are described. A non-stationary 
1$\times$2-dimensional model rendering possible an analytical
investigation of two-dimensional current-density and potential 
distributions transversal to the main current flow direction is derived
in section 3 for two cases characterized by different nonlinear electrical 
mechanisms. In section 4, stationary homogeneous states are determined
from the model equations and used to derive the current-voltage
characteristics for systems with uniform current-density distributions.
Finally, the stability of stationary homogeneous states is studied
in section 5, including a discussion of important parameter dependencies.

\section{Multilayer semiconductor structures as
1$\times$2-dimensional extended active media}

\begin{figure}
\centerline{{\epsfxsize=14cm \epsfbox{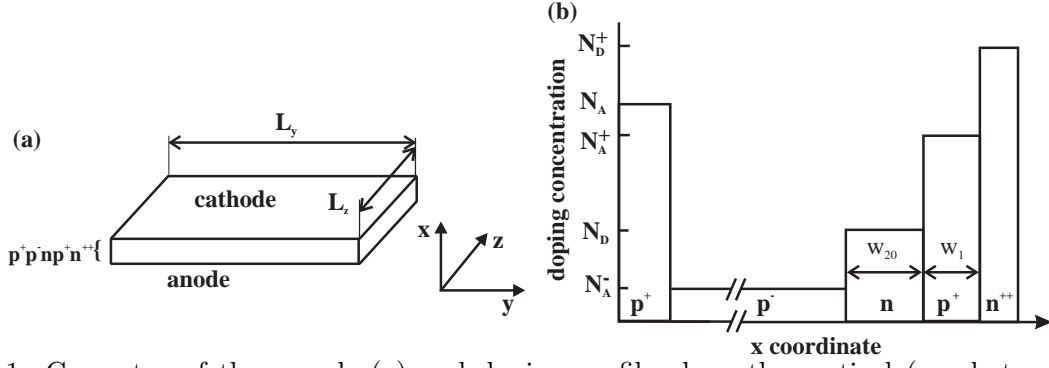}}}
\caption{Geometry of the sample (a) and 
	doping profile along the vertical
	(anode-to-cathode) direction (b).}
\label{fig1}
\end{figure}

The system under consideration consists of a
semiconductor structure with a vertical
width $w$ of several hundreds of $\mu$m
and with the two other (transversal)
widths $L_y, L_z \gg w $. 
Along the vertical direction the structure is characterized
by a $p^+$-$p^-$-$n$-$p^+$-$n^{++}$ doping profile as schematically
shown in Fig.~\ref{fig1}. The outer $p^+$ and $n^{++}$ layer of 
the structure are provided with metal contacts.
The internal design of the structure  possesses the following
peculiarities:

(i) Along the (vertical) $x$-direction 
the $p^+$-$p^-$-$n$-$p^+$-$n^{++}$ structure 
can be divided into an active (triggering) subsystem 
consisting of a four-layer thyristor-like $p^-$-$n$-$p^+$-$n^{++}$
structure and a subsystem of a distributed resistive layer
consisting essentially of the remainder $p^-$ layer. Note that the 
thyristor-like regeneration mechanism in the active subsystem
causes an autocatalytic increase 
of the current density in a certain current interval 
which is counteracted by the resistive layer.

The width of the $n$-$p^+$-$n^{++}$ part is only a few tens of $\mu$m
which is more than an order less than the whole width $w$ of the device.
The lifetime $\tau$ of the excess carriers in the bulk material is supposed
to be so small that conductivity modulation of the bulk due to the injection
of electrons and holes from the outer $n^{++}$ and $p^+$ layers
is prevented. This can be achieved by reducing the lifetime 
in the main part of the bulk in a certain distance to the anode contact 
leaving a relatively thin part of the $p^-$ layer with large $\tau$ 
nearby the inner $n$ layer.

(ii) The doping concentrations $N_A^+$ and $N_D^{+}$
in the layers forming the cathode emitter $p^+$-$n^{++}$ junction
[Fig.~\ref{fig1}(b)] are similar to those used 
in high frequency bipolar transistors.
The electron injection efficiency of this junction increases 
from nearly zero at very low current densities to a saturation value
of approximately 0.8-0.9 at rather small current densities 
of the order of 1 mA/cm$^2$.

(iii) The doping concentration $N_A^-$ of the $p^-$ layer
is much smaller than the doping concentration $N_D$ of the $n$ base.
Hence, for low currents, current transport through this junction is
dominated by a leakage current of electrons from the $n$ base into 
the $p^-$ emitter and, consequently, the injection efficiency of
the $p^-$-$n$ junction is very small.

(iv) The widths of the $p^+$ and $n$ base, $w_1$ and $w_{20}$, in 
the thyristor-like subsystem are much smaller than the 
diffusion lengths $L_n$ and $L_p$ of the minority
carriers in these layers:  $w_1 \ll L_n$, $w_{20} \ll L_p$.
Consequently, both base transport factors are 
about one. 

\begin{figure}
\centerline{{\epsfxsize=14cm \epsfbox{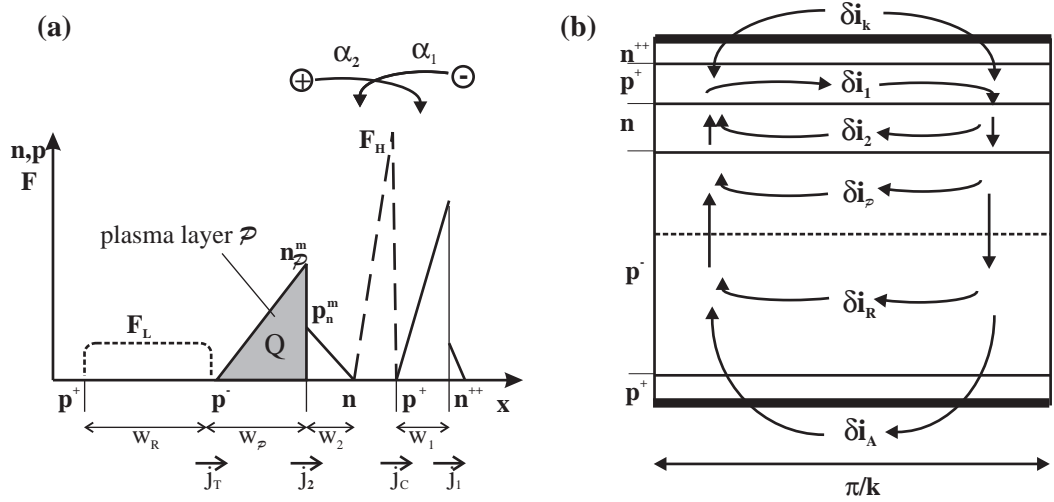}}}
\caption{Non-equilibrium plasma-field stratification
	along the vertical direction (a) and
	possible contours of small current-density fluctuations
	which are periodical along the transversal directions (b).}
\label{fig2}
\end{figure}

When a dc voltage is applied to the 
device such that the outer $n^{++}$ layer
is biased negatively with respect to the outer $p^+$ layer, the 
anode and cathode emitter junction are forward biased, while
the collector junction is reverse-biased. Thus, double injection
occurs inside the thyristor-like subsystem accompanied by a 
characteristic nonequilibrium plasma-field stratification
along the vertical direction $x$ as shown in Fig.~\ref{fig2}(a).
Due to the chosen doping proportion at 
the $p^-$-$n$ junction, which is just inverse to that in 
standard thyristors, there is a significant
electron leakage current from the $n$ base into the $p^-$ layer 
which obviously exceeds the electron injection current. Therefore, 
a rather large excess charge accumulates in a thin plasma layer
($\cal P$ layer).
If the carrier lifetime in this plasma layer is sufficiently large,
the local dynamics of this charge at non-stationary conditions
may be considered to be the most inertial process, thus
determining the temporal evolution of the whole device. The derivation
of the model equations in section 3 is based essentially on these 
physical properties.

The nonlinear feedback mechanism between the electron
and hole injection currents in the $p^-$-$n$-$p^+$-$n^{++}$
subsystem is similar to that in standard 
thyristors. However, there are remarkable peculiarities
and it is useful to distinguish between two cases, corresponding
to certain current intervals, which are characterized by
a particular nonlinear mechanism, respectively, causing 
the appearance of negative differential resistance in 
the current-voltage characteristic of the thyristor-like
subsystem. In analogy to the two-transistor approach for 
thyristors we 
consider the thyristor-like subsystem to be
composed of an $n$-$p^+$-$n^{++}$ and a $p^-$-$n$-$p^+$ 
transistor to specify the two different regenerative 
mechanisms. Current regeneration \cite{Ger1,Sze} in thyristors 
is possible for current intervals, in which the condition
$r_T^0 = dV_T/dj < 0$ holds, 
where $V_T$ and $j$ denote the thyristor
voltage and current, and $r_T^0$ is the differential resistance.
$dV_T/dj$ is implicitly given by the
steady state condition
\begin{eqnarray}
	\alpha_1(j,V_T) + \alpha_2(j,V_T) = 1,
	\label{eq:alpha_sum}                                    
\end{eqnarray}	
where $\alpha_1$ and
$\alpha_2$ denote the current gains of the $n$-$p^+$-$n^{++}$ 
and the $p^-$-$n$-$p^+$ transistor, respectively [Fig.~\ref{fig2}(a)].

Charge carrier recombination in the space
charge region of the emitter junction of a transistor influences
the current gain significantly \cite{Ger1,Sze} and typically leads to
a monotonic increase of the current gain with increasing current
for not too large currents.

Consider the device parameters to be of such kind
that the current gain $\alpha_1$ approaches a value close
to one at relatively low currents ($j < 10^{-3} $A/cm$^2$). 
Then Eq.~(\ref{eq:alpha_sum}) can be fulfilled at a very low
injection efficiency of the $p^-$-$n$ emitter junction.
Thus, this case -- subsequently called {\em low current range} -- 
is roughly characterized by
$0 < \alpha_1 = \alpha_1(j) \le 1$ and 
$0 < \alpha_2 \approx const \ll 1$.

If the doping profile and other parameters are chosen such
that $\alpha_1$ saturates in the low current range
before Eq.~(\ref{eq:alpha_sum}) is true, regeneration does
not occur at low currents. Nevertheless, at larger currents 
regeneration might become possible again when
$\alpha_2$ increases monotonously with the current. This
can occur at sufficiently large current, when a plasma layer
$\cal P$ evolves in the $p^-$ layer near the $p^-$-$n$ junction 
and the injection of excess holes from this plasma layer into the
$n$ base increases superlinearly as proposed by Fletcher
\cite{Fle57}. Consequently, the $\cal P$ layer plays the role
of a plasma emitter, the efficiency of which rises strongly with
the current so that Eq.~(\ref{eq:alpha_sum}) can be fulfilled.
This case is called {\em moderate current range} subsequently
and is roughly characterized by
$0 < \alpha_1 \approx const < 1$ and 
$0 < \alpha_2 = \alpha_2(j) < 1$.

Before we start with a detailed analysis of these two cases
let us turn to a short discussion on important transversal processes
which have to be taken into account for an analytic description, too.
To this end, consider a  
so-called differential element of the whole 
$p^+$-$p^-$-$n$-$p^+$-$n^{++}$ structure,
which is small in $y$ and $z$ direction.
Fig.~\ref{fig2}(b) shows schematically contours of possible current
fluctuations  in such an element which may lead to a destabilization 
of the homogeneous current flow. Obviously,
there are three channels along which transversal currents 
effect a transversal spatial coupling in the
thyristor-like subsystem: the $p^+$ base, the $n$ base, and
the possibly conductivity modulated $\cal P$ layer
in the $p^-$ bulk, in which
the corresponding transversal fluctuations $\delta i_1$, $\delta i_2$,
and $\delta i_{\cal P}$ may arise.
For the resistor subsystem transversal current fluctuations
$\delta i_R$ are taken into account; they are distributed
over the whole bulk due to three-dimensional deformations of the
potential distribution $\varphi(x,y,z,t)$; these deformations are
caused by potential fluctuations at the interface between the
two subsystems, which in turn result from local current fluctuations 
in the thyristor-like subsystem.

The metallic layers forming the anode and cathode contact on the
top and bottom of the device represent further channels denoted
by $\delta i_A$ and $\delta i_K$ in Fig.~\ref{fig2}(b) and render possible
the formation of closed loops of current fluctuations inside
the system without any interaction with elements of the external circuit.
Thus, all differential elements of the system are
connected to each other by both the vertical feedback mechanism
and transversal couplings which have to be considered 
in a self-consistent description as outlined in detail in
the following sections.

%\newpage
\section{Non-stationary (1$\times$2)-dimensional analytical model}

\subsection{The thyristor-like subsystem}

As mentioned above the excess plasma of the $p^{-}$ layer
is considered to be the most inertial part of the system.
Consequently, all other nonstationary transport effects are
neglected. We start with a derivation of the basic equations
for the low current range. Based on these results the 
peculiarities concerning the case of moderate currents are discussed
in a second subsection.

\subsubsection{Low current range}
In the low current case the dynamics of the plasma emitter
can be described in terms of the excess charge density
$Q = Q(y,z,t)$ per unit square in the plasma layer $\cal P$:
\begin{eqnarray}
\label{eq:Q'-dgl}                                       %Q'-dgl
  \frac{\partial Q}{\partial t} =
    -\frac{Q}{\tau} + j_{2n} - j_{Tn} +
    \mbox{\boldmath $\nabla_{\perp} \cdot i_{{\cal P} n}$},
\end{eqnarray}
where
\begin{eqnarray}
\label{eq:i_pl_low}                                       %Q'-dgl
    \mbox{\boldmath $i_{{\cal P} n}$} = 
    	-b(<n>/N_A^-)\sigma_{\cal P} w_{\cal P} 
    	\mbox{\boldmath $\nabla_{\perp}$} V_T 
        + D_n \mbox{\boldmath $\nabla_{\perp}$} Q
    \approx D_n \mbox{\boldmath $\nabla_{\perp}$}Q 
\end{eqnarray}
is the sheet electron current-density in the plasma layer,
$\mbox{\boldmath $\nabla_{\perp}$} = 
(\partial / \partial y, \partial / \partial z)$ 
the two-dimensional $\mbox{\boldmath $\nabla$}$-operator,
$D_n$ the diffusion coefficient of electrons,
$\tau$ the lifetime of excess charge carriers at low injection,
$j_{2n}$ the $x$ component of the electron current-density
at the boundary between the plasma layer $\cal P$
and the $n$ base,
$j_{Tn}$ the $x$ component of the electron current-density
leaving the thyristor-like subsystem and entering the
resistor subsystem.\footnote{Note that all quantities denoted by 
$j_{...}$ refer to the $x$ component of the corresponding
current-density vector; for a better readability the supplement
{\em $x$ component} is omitted in the following explanation.
Accordingly, quantities denoted by 
$\mbox{\boldmath $i_{...}$} = (i_{...,y},i_{...,z})$ 
refer to pure sheet current densities per unit length
in a certain layer of the structure.
Physical constants as well as material and design parameters are listed 
separately in Tables~\ref{table1} and \ref{table2}.}
$V_T$ denotes the voltage drop across the thyristor-like subsystem, 
$N_A^-$ the acceptor concentration of the $p^-$ layer,
$\sigma_{\cal P}$ the conductivity of the plasma layer $\cal{P}$,
which is equal to $q \mu_p N_A^-$ in the low current range,
$q$ the elementary charge,
$b = \mu_n / \mu_p$,
$\mu_n$ and $\mu_p$ the electron and hole mobility, and
$<n> = \int\limits_{0}^{w_{\cal{P}}} n(x) dx$ the average electron density
per unit square in the plasma layer with width $w_{\cal{P}}$. 

Following the charge-control model of transistors 
and thyristors (see, e.g., Ref.~\cite{Ger1}) 
the variable $j_{Tn}$ can be expressed by
\begin{eqnarray}
\label{eq:tet'}                                     	 %tet'
  j_{Tn}=\frac{Q}{\theta}, \qquad
  \theta=\frac{w_{\cal P}^2}{2D_n}.
\end{eqnarray}
Thus, the charge equation can be re-written as
\begin{eqnarray}
\label{eq:Char'-dgl}					%Char'-dgl
 \frac{\partial Q}{\partial t} =
    D_n\Delta_{\perp} Q + j_{2n} - \frac{Q}{\tau_*},                               
\end{eqnarray}
with
$1/ \tau_* = 1/ \tau + 1 / \theta$.
The variable $Q$ can be further approximated in the following
way by means of 
the maximum value of the excess charge carrier concentration
$n_{\cal P}^m$ in the plasma layer:
\begin{eqnarray}
\label{eq:approx}                                      	%approx
    Q\simeq \frac{qw_{\cal P}n_{\cal P}^m}{2}.
\end{eqnarray}
Relations connecting the concentrations of the minority charge carriers
at both sides of the $p^-$-$n$ junction, $n_{\cal P}^m$ and $p_n^m$,
and the hole current-density $j_{2p}$ at the interface between the 
plasma layer and the $n$ base are provided by
the classical theory of $p$-$n$ junctions 
(see, e.\,g., Ref.~\cite{Sze}):
\begin{eqnarray}
\label{eq:rel'}                                      %rel'
  n_{\cal P}^m N_A^-&=&p_n^m N_D,\\
\label{eq:rel'1}				     %rel'1
  j_{Cp}=j_{2p}&=&\frac{q D_p p_n^m}{w_2(V_C)},
\end{eqnarray}
where 
$N_D$ is the donor concentration in the $n$ base,
$w_2$ the effective width of the $n$ base,
which depends on the voltage drop $V_C$ across
the $n$-$p^+$ collector junction, and
$D_p$ the diffusion coefficient of holes.
Note that due to the assumption $w_2 \ll L_p$ [Sect. 2, assumption (iv)]
the hole current-density $j_{Cp}$ at the collector is equal to $j_{2p}$.

The dependence of the base width $w_2$ on the collector voltage 
drop $V_C(y,z,t)$ can be approximated by 
(see Ref.~\cite{Sze}):
\begin{eqnarray}
\label{eq:U'}                                         %U'
  w_2(V_C)&=&w_{20}\left[1-\left(\frac{V_C}{V_{pth}}\right)^{1/2}\right],\\
  {\mathrm with} \hspace{2em}
  V_{pth}&=&\frac{qN_D w_{20}^2}{2\epsilon \epsilon_0},
  \nonumber
\end{eqnarray}
where 
$w_{20}$ denotes the whole width of the $n$ base,
$V_{pth}$ the punch-through voltage of the $n$-$p^+$ junction,
$\epsilon_0$ and $\epsilon$ the vacuum permittivity
and the dielectric constant of silicon.

Equations~(\ref{eq:approx}) - (\ref{eq:rel'1})
yield an expression linking the hole current-density 
at the collector junction, $j_{Cp} = j_{2p}$, 
the charge $Q$ and the collector voltage $V_c$:
\begin{eqnarray}
\label{eq:QJc'}                                           %QJc'
    Q &=& \tau'\left(1-\sqrt{\frac{V_C}{V_{pth}}}\right)j_{Cp},\\
    {\mathrm with} \hspace{2em}
   (\tau')^{-1}&=&\frac{ N_A^-}{N_D} \frac{2D_p}{w_{\cal P} w_{20}}.
        \nonumber %
\end{eqnarray}

In the $n$-$p^+$-$n^{++}$ transistor the local current 
balance for a differential element of the $p^+$ base is given by:
\begin{eqnarray}
\label{eq:div1'}
  j_C-j_1     =   \mbox{\boldmath $\nabla_{\perp} \cdot i_1$}
             &=&  -\sigma_1 w_1 \Delta_{\perp}\varphi_1, \\
    {\mathrm with} \hspace{2em}
    \sigma_1 &=&  q\mu_{p}N_A^+ ,  \nonumber                   %div1'
\end{eqnarray}
where
$j_C$ and $j_1$ denote the total current-densities at the
$n$-$p^+$ collector and $p^+$-$n^{++}$ emitter junction,
$\sigma_1$ the conductivity of the $p^+$ base,
and
$\varphi_1$ the voltage drop across the emitter junction.

Because of $w_1 \ll L_n$, the electron current-density at the collector
$j_{Cn}=\alpha_1 j_1$ is practically equal to the electron
current-density $j_{1n}$ injected by the $n^{++}$ layer.
Nevertheless, since the hole leakage from the $p^+$ base
into the $n^{++}$ emitter is strongly nonlinear 
we have to take into account a $j$-dependence
(or bias-dependence) of the parameter $\alpha_1$.
The parametric dependence between the 
total emitter current-density $j_1$
and its electron injection component $j_{1n}$ 
is given as an implicit function 
coupling the emitter voltage $\varphi_1$ with the
emitter current-density $j_1$:
\begin{eqnarray}                                        %a'
\label{eq:a'}
  0=F(j_1, \varphi)&=&\left(j_{1ns} + j_{1ps}\right) 
  		  \exp{\left(q\varphi_1/kT\right)}
                  +j_{1Rs} \exp{\left(A \varphi_1\right)} - j_{1},\\%
  j_{Cn} \approx j_{1n}&=&j_{1ns} \exp{\left(q\varphi_1/kT\right)}.%
\end{eqnarray}
Here $j_{1ns}$, $j_{1ps}$ and $j_{1Rs}$ are saturation
values of the electron injection current-density, and
the linear and nonlinear hole leakage
current-densities, respectively.
The parameter $A$ depends on the concrete mechanism of the
hole leakage in the $p^+$-$n^{++}$ junction \cite{GN1}.
For the case under study we assume $A=q/2kT$, i.\,e.
Sah-Noyce-Shockley recombination in the space charge region.

In the $\cal P$-$n$-$p^+$ transistor the local current balance for
a differential element of the $n$ base
and of the plasma layer $\cal P$ is given by:
\begin{eqnarray}
  j_2-j_C&=&\mbox{\boldmath $\nabla_{\perp} \cdot i_2$} 
  						  \nonumber \\
	 &=& j_{2n} - j_{Cn} 			  \nonumber \\
\label{eq:div2'}					%div2'
  	 &=& -\mbox{\boldmath $\nabla_{\perp}$}
  	      \left[\sigma_2 w_2(V_C)
              \mbox{\boldmath $\nabla_{\perp}$} 
                   (V_C + \varphi_1)\right], \\
  j_T-j_2&=&\mbox{\boldmath $\nabla_{\perp} \cdot i_{\cal P}$} 
  							\nonumber \\
\label{eq:div2'a}					%div2'a
	 &=& -\sigma_{\cal P} w_{\cal P}\Delta_{\perp} V_T +
             D_n \Delta_{\perp} Q, \\
\end{eqnarray}
where $\sigma_2 = q\mu_n N_D$ 
is the conductivity of the $n$ base.
Note, that low injection conditions are assumed for the plasma emitter.

Combining Eqs. (\ref{eq:div2'}) and (\ref{eq:div2'a}) yields:
\begin{eqnarray}
  j_C   &=& j_T 
            -	\mbox{\boldmath $\nabla_{\perp} \cdot i_{\cal P} $}
            -	\mbox{\boldmath $\nabla_{\perp} \cdot i_2 $} \nonumber \\
\label{eq:P_em'}
        &=& j_T
            +	\mbox{\boldmath $\nabla_{\perp}$}\left[\sigma_2 w_2(V_C) 
      		\mbox{\boldmath $\nabla_{\perp}$}(V_C + \varphi_1)\right]
      	    +	\sigma_{\cal P} w_{\cal P} \Delta_{\perp} V_T
      	    -	D_n\Delta_{\perp}Q,\\
  j_{Cp}&=& j_C-j_{Cn}.                                    %P_em'
\end{eqnarray}

The voltage drop $V_T$ across the thyristor-like subsystem is given by
$V_T = V_C + \varphi_1 + \varphi_2$, where $\varphi_2$ is the
voltage drop across the $p^-$-$n$ junction. 
Using $p_n^m = p_{n0}\exp{(q \varphi_2/kT)}
             = n_{p0}(N_A^-/N_D)\exp{(q \varphi_2/kT)}$ 
and Eqs.~(\ref{eq:approx}) and (\ref{eq:rel'}) the following relation between
$Q$ and $\varphi_2$ can be derived:
\begin{eqnarray}
  Q &=& \frac{q w_{\cal P} n_{p0}}{2} \exp{(q \varphi_2/kT)} \nonumber \\
\label{eq:phi_2}					%phi_2
    &=& \frac{q w_{\cal P}}{2} \frac{n_i^2}{N_A^-} \exp{(q \varphi_2/kT)},
\end{eqnarray}
where $p_{n0}$ and $n_{p0}$ denote the equilibrium minority carrier
concentrations of the $n$ base and $p^-$ bulk, respectively, and 
$n_i$ is the intrinsic carrier concentration.

\subsubsection{Moderate current range}

Now we indicate the peculiarities
occuring at moderate currents, which case is characterized by
a high injection level in the $p^-$ bulk.
In this case ambipolar effects
in the plasma layer should be included into
consideration. In the frame of the charge-control model the
following relations between the
current densities $j_{Tn}$ and $j_T$, 
the sheet current densities 
$\mbox{\boldmath $i_{{\cal P} n}$}$ and 
$\mbox{\boldmath $i_{\cal P}$}
 	= \mbox{\boldmath $i_{{\cal P} n}$}
        + \mbox{\boldmath $i_{{\cal P} p}$} $
and the charge $Q$ in the plasma layer $\cal P$ are valid:
\begin{eqnarray}
\label{eq:tet1}							%tet1
  j_{Tn} &=& \frac{b}{b+1}j_T+\frac{Q}{\theta_h}, \\       	
\label{eq:tet2}							%tet2
  \mbox{\boldmath $i_{\cal P}$} 
	&=&  \mbox{\boldmath $i_{{\cal P} n}$} 
	 +   \mbox{\boldmath $i_{{\cal P} p}$}
	 =   -w_{\cal P}\sigma_{\cal P}(Q)
	              \mbox{\boldmath $\nabla_{\perp}$} V_T 
	 +   (b-1)D_p \mbox{\boldmath $\nabla_{\perp}$} Q,\\
\label{eq:tet3}							%tet3
  \mbox{\boldmath $i_{{\cal P} n}$} 
  	&=&  -\frac{b}{b+1}w_{\cal P}\sigma_{\cal P}(Q)
  	          \mbox{\boldmath $\nabla_{\perp}$} V_T
  	 +   bD_p \mbox{\boldmath $\nabla_{\perp}$} Q.
\end{eqnarray}
Here $w_{\cal P}\sigma_{\cal P}(Q)=(b+1)\mu_p Q$ is the
transversal conductance of the plasma layer, 
$\theta_h = w_{\cal P}^2/(2D_h)$ the transit time through
the plasma layer, and
$D_h = 2bD_p/(b+1)$ the ambipolar diffusion coefficient.
Note that Eqs.~(\ref{eq:tet1}) and (\ref{eq:tet3}) 
substitute Eqs.~(\ref{eq:tet'}) and (\ref{eq:i_pl_low}),
respectively.

The charge balance equation for the plasma layer
should be re-written as
\begin{eqnarray}
\label{eq:char-dgl}
    \frac{\partial{Q}}{\partial{t}} 
      =  -\frac{Q}{\tau_{h}}+j_{2n} - j_{Tn}
      +  \mbox{\boldmath $\nabla_{\perp} \cdot i_{{\cal P} n}$},
\end{eqnarray}
where $\tau_h$ is the high-injection lifetime. 
By inserting Eqs.~(\ref{eq:tet1}) and (\ref{eq:tet3}) we obtain
\begin{eqnarray}
\label{eq:char-high}
    \frac{\partial{Q}}{\partial{t}} =
       bD_p\Delta_{\perp}Q-
        \frac{Q}{\tau_h}-\frac{Q}{\theta_h} +
        j_{2n} - \frac{b}{b+1}j_T -
        \frac{b}{b+1} \mbox{\boldmath $\nabla_{\perp}$}
        \left[w_{\cal P} \sigma_{\cal P}(Q) 
        \mbox{\boldmath $\nabla_{\perp}$} V_T \right]
\end{eqnarray}
[compare with Eq.~(\ref{eq:Char'-dgl})].

The relation between the excess carrier concentrations
at the borders of the space charge region of the $p^-$-$n$ junction
changes from Shockley's to Fletcher's \cite{Fle57} form:
\begin{eqnarray}
\label{eq:rel}
  (n_{\cal P}^m)^2=p_n^m N_D                     		%rel
\end{eqnarray}
[compare with Eq.~(\ref{eq:rel'})].

Taking into account Eqs.~(\ref{eq:approx}), (\ref{eq:rel'1}), 
and (\ref{eq:rel}) 
we obtain in analogy to Eq.~(\ref{eq:QJc'})
an expression linking the charge $Q$ in the plasma layer,
the voltage drop $V_T \approx V_C$ across the 
$\cal P$-$n$-$p^+$-$n^{++}$ subsystem and 
the hole current-density $j_{Cp}$ of the collector:
\begin{eqnarray}
\label{eq:QJc}							%QJc
  	Q  =  Q_m\left[\frac{w_2(V_C)}{w_{20}}\right]^{1/2}
	      \left[\frac{j_{Cp}}{(1-\alpha_1)j_m}\right]^{1/2}, 
\end{eqnarray}
where the normalizing quantities $Q_m$ and $j_m$ are connected by
\begin{eqnarray}
\label{eq:Q_m}							%Q_m
	Q_m = \left[\frac{qN_D w_{\cal P}^2w_{20}(1-\alpha_1)j_m}      
	      {4D_p}\right]^{1/2}. 
\end{eqnarray}

As already mentioned above, the current gain $\alpha_1$ of the
$n$-$p^+$-$n^{++}$ transistor is approximately current-independent in 
the moderate current range.
Then,  the  electron current-density $j_{Cn}$ of the collector 
is given by
\begin{eqnarray}                                         %jCn
\label{eq:jCn}
  j_{Cn}=\alpha_1 j_1, 
  \hspace{2em} {\mathrm with \hspace{1em}} 
  \alpha_1 = const > \frac{b}{b+1}.
\end{eqnarray}

In the $\cal P$-$n$-$p^+$ transistor 
conductivity modulation in the plasma layer $\cal P$ due to excess
carrier concentrations should be taken into account 
at high injection level; consequently Eq.~(\ref{eq:div2'a})
is substituted by:
\begin{eqnarray}
  j_T-j_2 &=& \mbox{\boldmath $\nabla_{\perp} \cdot 
                   i_{\cal P}$} \nonumber \\
\label{eq:div2}                                      %div2
	  &=& (b-1)D_p\Delta_{\perp} Q
           -  \mbox{\boldmath $\nabla_{\perp}$}
              \left[w_{\cal P} \sigma_{\cal P}(Q) 
              \mbox{\boldmath $\nabla_{\perp}$} V_T \right].
\end{eqnarray}

On account of vanishing field and diffusion currents at the
transversal borders of the sample ($y = 0,L_y; z=0,L_z$)
the following boundary conditions for
the $\cal P$-$n$-$p^+$-$n^{++}$ subsystem have been chosen for
both the low-injection and the high-injection case:
\begin{eqnarray}                                      %T_conds
\label{eq:T_conds}
    (\mbox{\boldmath $n \cdot \nabla_{\perp}$} Q)
        \vert_{y=0,L_y;~ z=0,L_z} &=&0, \\
    (\mbox{\boldmath $n \cdot \nabla_{\perp}$} \varphi_1)
        \vert_{y=0,L_y;~ z=0,L_z} &=&0, \\
    (\mbox{\boldmath $n \cdot \nabla_{\perp}$} \varphi_2)
        \vert_{y=0,L_y;~ z=0,L_z} &=&0, \\
\label{eq:T_conds4}					%T_conds4
    (\mbox{\boldmath $n \cdot \nabla_{\perp}$} V_T)
        \vert_{y=0,L_y;~ z=0,L_z} &=&0.
\end{eqnarray}
Here ${\bf n}$ is a unit vector normal to the transversal
surfaces of the sample.

\subsection{Resistor-like subsystem and external circuit}
In the resistor-like part of the $p^-$ bulk displacement currents
are neglected. The three-dimensional potential distribution
in this region is described by Laplace's equation:
\begin{eqnarray}
\label{eq:Lapl-dgl}                                     %Lapl-dgl
           \Delta \varphi=0 %
\end{eqnarray}
with the boundary conditions
\begin{eqnarray}
\label{eq:condphi}                               %condphi
  \varphi_{\vert x=0} =V_A,\qquad
        \varphi\vert_{x=w_R} =V_T,\\
\label{eq:condphi_2}                             %condphi_2
    (\mbox{\boldmath $n \cdot \nabla_{\perp}$} \varphi)    
    \vert_{y=0,L_y;~ z=0,L_z}~=~0,
\end{eqnarray}
where 
$V_A$ denotes the voltage drop across the whole sample,
$w_R$ the width of the resistive layer, and
$\Delta = \partial^2/\partial x^2 +
        \partial^2/\partial y^2 +
        \partial^2/\partial z^2 $
the three-dimensional Laplace-operator.

The current-density vector 
$\mbox{\boldmath ${j_R}$}(x,y,z,t)$
inside the $p^-$ bulk is connected with the local
potential $\varphi(x,y,z,t)$ according to
\begin{eqnarray}
\label{eq:Ohm}                                         %Ohm
  \mbox{\boldmath ${j_R}$} 
  = -\sigma_R \mbox{\boldmath $\nabla$} \varphi, %
\end{eqnarray} %
and its $x$ component $j_{R,x}$ at the boundary $x=w_R$
"feeds" the thyristor-like subsystem locally so that
\begin{eqnarray}                                            %j_T
\label{eq:j_T}
  j_{R,x}|_{x=w_R} = j_T (y,z).
\end{eqnarray}

The $p^+$-$p^-$-$n$-$p^+$-$n^{++}$ sample is driven by
an ideal voltage source $V_S$ via an external 
load resistor $R_L$. This yields the following
load line equation:
\begin{eqnarray}                                            %load
\label{eq:load}
   V_A(t) = V_S - R_L\int_{L_y,L_z}{j_T(y,z,t)~dydz}.
\end{eqnarray}

The sets of equations 
(\ref{eq:Q'-dgl}) - (\ref{eq:phi_2}),
(\ref{eq:T_conds}) - (\ref{eq:load}) 
and 
(\ref{eq:U'}), (\ref{eq:div1'}), (\ref{eq:div2'}),
(\ref{eq:phi_2}) - (\ref{eq:load}) 
present a self-consistent non-stationary $1\times2$-dimensional
description of the system under consideration for
the low and moderate current range, respectively.

%\newpage
\section{Homogeneous stationary states and corresponding
	current-density vs. voltage characteristics}

To describe the homogeneous stationary states
Eqs.~(\ref{eq:Char'-dgl}), (\ref{eq:div1'}),
(\ref{eq:div2'}), (\ref{eq:P_em'}), and (\ref{eq:Lapl-dgl})
for the low current range and
Eqs.~(\ref{eq:div1'}), (\ref{eq:div2'}),
(\ref{eq:char-high}), (\ref{eq:div2}), and (\ref{eq:Lapl-dgl})
for the moderate current range 
with all time and space derivatives set to zero have to be solved:
   $\partial Q/\partial t                      = 0, 
   \mbox{\boldmath $\nabla_{\perp}$} Q         = 0,
   \mbox{\boldmath $\nabla_{\perp}$} \varphi_1 = 0,
   \mbox{\boldmath $\nabla_{\perp}$} \varphi_2 = 0,
   \mbox{\boldmath $\nabla_{\perp}$} \varphi   = 0,
   \mbox{\boldmath $\nabla_{\perp}$} V_C       = 0,
   \mbox{\boldmath $\nabla_{\perp}$} V_T       = 0$.
This leads to $j_T = j_2 = j_C = j_1 = j = const$.

\subsection{Low current range}   \label{sec:hom_lil}

The current gain $\alpha_1$ of the $n$-$p^+$-$n^{++}$ transistor
can be written as
\begin{eqnarray}
\label{eq:al}                                           % al
    \alpha_1 =  \alpha_1(\varphi_1)
    	     =  \frac{j_{1n}}{j}
             =  \frac{j_{1ns}\exp{(q\varphi_1/kT)}}
                     {(j_{1ns}+j_{1ps})\exp{(q\varphi_1/kT)}+
                     j_{1Rs}\exp{(q\varphi_1/2kT)}}.
\end{eqnarray}
Combining the charge balance equation (\ref{eq:Char'-dgl})
for the stationary homogeneous state 
with $j_{2n} = j_{Cn} = \alpha_1 j$ and Eq.~(\ref{eq:al}) 
yields
\begin{eqnarray}                                        %stat'
\label{eq:stat'}
  Q=\tau_* \alpha_1 j=\tau_*j_{1ns} \exp{(q\varphi_1/kT)},
 	{\mathrm with} \hspace{2em}
        \tau_*^{-1}=\tau^{-1}+\theta^{-1}.
\end{eqnarray}
According to Eq.~(\ref{eq:QJc'}) the variable $Q$
is proportional to $j_{Cp}$, which is equal to $(1-\alpha_1)j$;
together with Eq.~(\ref{eq:U'})
and $V_T \approx V_C$ one obtains:
\begin{eqnarray}                                        %statQ'
\label{eq:statQ'}
  Q= \tau' \left[1-\left(\frac{V_T}{V_{pth}} \right)^{1/2}\right]
     \left(1-\alpha_1 \right) j.
\end{eqnarray}
Combining Eqs.~(\ref{eq:stat'}) and (\ref{eq:statQ'})
leads to the following parametrically determined
$j(V_T)$-characteristic of the thyristor-like
$\cal P$-$n$-$p^+$-$n^{++}$ subsystem 
for the low current range:
\begin{eqnarray}                                         %IV'
\label{eq:IV'}
  V_T=V_{pth}~\left[1-\frac{\tau_*}{\tau'}
    \frac{\alpha_1 (\varphi_1)}
    {\left[1-\alpha_1(\varphi_1)\right]}\right]^2, 
\end{eqnarray}
where the voltage drop $\varphi_1$ across the
$p^+$-$n^{++}$ emitter and the current density $j$
are coupled by
\begin{eqnarray}                                         %IV''
\label{eq:IV''}
  j=\left(j_{1ns} + j_{1ps}\right)\exp{(q\varphi_1/kT)}
                  + j_{1Rs}\exp{(q \varphi_1/2kT)}.
\end{eqnarray}

The differential resistance of the thyristor-like 
subsystem for the homogeneous state under low current
conditions is then given by:
\begin{eqnarray}
\label{eq:Rd'}                                         %Rd'
  r_T^0      &=& \frac{dV_T}{dj} = \frac{dV_T}{d\alpha_1}
               \frac{d\alpha_1}{d\varphi_1} 
               \frac{d\varphi_1}{dj},\\
  \frac{d\varphi_1}{dj} 
           &=& -\frac{\partial F(j,\varphi_1)}{\partial j}
               \times\left[\frac{\partial F(j,\varphi_1)}
               {\partial \varphi_1} \right]^{-1} . \nonumber
\end{eqnarray}

\subsection{Moderate current range}

Using the charge balance equation 
(\ref{eq:char-high}) and 
$j_{2n} = j - j_{Cp} = \alpha_1 j$
we obtain the following relation between $Q$ and $j$:
\begin{eqnarray}
\label{eq:stat}                                        %stat
        	Q   &=& \tau_* g_0 j, \\
        {\mathrm with}  \hspace{2em}
        \tau_*^{-1} &=& \tau_h^{-1}+\theta_h^{-1},
       		g_0  = \alpha_1 - \frac{b}{b+1} \nonumber.
\end{eqnarray}
On the other hand, for the homogeneous case
the variable $Q$ depends on $j$ according to the
relation 
\begin{eqnarray}                                          %statQ
\label{eq:statQ}
  Q=Q_m \left[1-\left( \frac{V_T}{V_{pth}}\right)^{1/2}\right]^{1/2}
    \left(\frac{j}{j_m}\right)^{1/2},
\end{eqnarray}
which follows from Eqs.~(\ref{eq:U'}),
(\ref{eq:QJc}) together with 
$j_{Cp} = (1-\alpha_1)j$ and $V_T \approx V_C$.

Choosing 
\begin{eqnarray}
\label{eq:jm}                                          %jm
  j_m=\frac{qN_D w_{\cal P}^2 w_{20} (1-\alpha_1)}{4D_p\tau_*^2 g_0^2}
\end{eqnarray}
and equalizing Eqs.~(\ref{eq:stat}) and 
(\ref{eq:statQ}) yields together with 
Eq.~(\ref{eq:Q_m}) the following expression for the
$j(V_T)$-characteristic of the thyristor-like subsystem
for the case of moderate currents:
\begin{eqnarray}
\label{eq:IV}                                           %IV
    j&=&j_m \left[1-\left( \frac{V_T}{V_{pth}}\right)^{1/2}\right],\\
\end{eqnarray}
where $j_m$ and $V_{pth}$ are given
by Eqs.~(\ref{eq:jm}) and (\ref{eq:U'}).

The specific differential resistance of the thyristor-like 
subsystem in the homogeneous state for this case is:
\begin{eqnarray}
\label{eq:r_T}                                   %r_T
  r_T^0  &=& \frac{dV_T}{dj} = - r_T^m\left(1-\frac{j}{j_m}\right)
          = - r_T^m\left(\frac{V_T}{V_{pth}}\right)^{1/2}, \\  
    {\mathrm with}  \hspace{2em} 
  r_T^m  &=& \frac{2V_{pth}}{j_m} \nonumber.
\end{eqnarray}

\subsection{The global current-density vs. voltage characteristic}
The global $j(V_A)$-characteristic of the complete 
$p^+$-$p^-$-$n$-$p^+$-$n^{++}$ structure ensues from
\begin{eqnarray}
\label{eq:Volt'}                                            %Volt'
  V_A(j)=V_T+V_R,
\end{eqnarray}
where $V_R=r_R^0 j$ denotes the voltage drop across the
resistive layer, the resistance of which is determined
by $r_R^0=w_R/(q \mu_p N_A^-)$.

\subsection{Calculated current-density vs.~voltage characteristics}

\begin{figure}
\centerline{{\epsfxsize=9cm \epsfbox{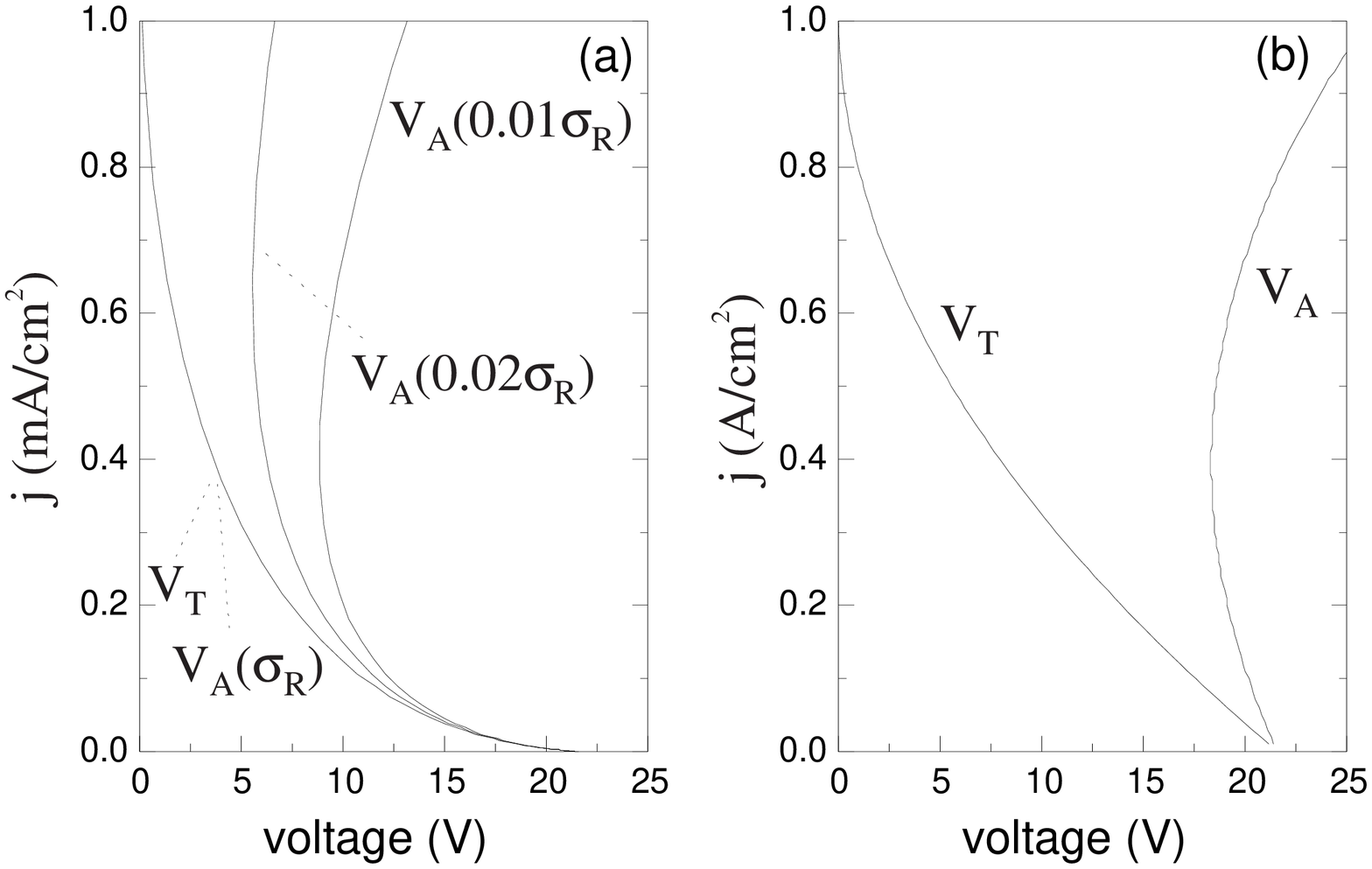}}}
\caption{Current-density vs. voltage characteristics  
	for the low (a) and moderate current range (b).
	$V_A$ and $V_T$ denote the bias of the total system and 
	the thyristor-like subsystem, respectively. 
	Parameter values for (a)
	are indicated in Table \protect{\ref{table2}}, 
	modified parameters for (b) are $w_R = 2 \cdot 10^{-2}$ cm and 
	$\alpha_1 = j_{1ns}/(j_{1ns}+j_{1ps})$ = 0.821.}
\label{fig3}
\end{figure}

By properly adjusting the design parameters of the $n^{++}$ layer 
it is possible to realize the appearance of a negative differential
resistance for both the low and moderate current case. In the calculations
presented below the saturation current-densities $j_{1ns}$ and $j_{1ps}$ 
have been used to adjust the two cases.
Figure \ref{fig3}(a) shows stationary current-density 
vs.~voltage characteristics for the low current case. 
Besides the characteristic $j(V_T)$ of the thyristor-like subsystem
and the characteristic of the total structure $j(V_A)$,
which are both calculated for the set of
standard parameters, two further $j(V_A)$-characteristics for
reduced values of the conductivity of the $p^-$ bulk are plotted.
For standard  parameter values the bulk resistance of the
$p^-$ bulk is not sufficiently large to compensate the negative
differential resistance of the thyristor-like subsystem at low currents.
For this reason the two curves $j(V_T)$ and $j(V_A)$ are nearly 
identical and cannot be distinguished in the diagram. In fact,
the current-density $j_{NDR}$, at which the negative differential resistance
transforms to a positive one, are nearly equal and approximately
1 mA/cm$^2$ for the two cases. 
$j_{NDR}$ can be reduced by a considerable amount only,
if the resistance of the $p^-$ bulk is enlarged at least by a factor of
ten, as illustrated by the two other curves in Fig.~\ref{fig3}(a).

Figure \ref{fig3}(b) shows current-density vs.~voltage characteristics
for the case of moderate currents and the standard value of the
conductivity of the $p^-$ bulk. 
The current-density $j_{NDR}$, at which the differential
resistance changes from negative to positive values, is 0.45 A/cm$^2$
for the total structure, i.\,e. considerably lower than the corresponding
value ($\approx $ 1 A/cm$^2$) for the thyristor-like subsystem. 

The blocking voltage is approximately equal to the punch-through 
voltage $V_{pth}$ for both the low and moderate current case. 
This is because the
emitter efficiencies are very small at low currents, so that a
hole injection current from the anode emitter, that is sufficiently 
large to initiate the regenerative process, becomes possible only,
when the effective width $w_2(V_C)$ of the $n$ base is 
approximately zero, i.\,e. under punch-through condition.

%\newpage
\section{Transversal stability analysis} 
In this section we analyze the stability of 
stationary homogeneous states along the transversal dimensions
with respect to small 1$\times$2-dimensional fluctuations of
the space-dependent variables ${\bf Y}$.
The standard approach for the linearization procedure implies 
to make the following Ansatz
for all variables in the vicinity of a stationary state
${\bf Y}_{stat}$
\begin{eqnarray}
\label{eq:linear}                                 % linear
  {\bf Y}(x,y,z,t)={\bf Y}_{stat} +
               \delta{\bf Y}(x,y,z)\exp{(\zeta t)}.
\end{eqnarray}

The small variations $\delta{\bf Y}(x,y,z)$ introduced here
have to satisfy the same boundary conditions as the variables
${\bf Y}(x,y,z,t)$ 
[see Eqs.~(\ref{eq:T_conds})-(\ref{eq:T_conds4}) and 
	  (\ref{eq:condphi}), (\ref{eq:condphi_2})].

\subsection{Fluctuations of the potential in the resistive layer}
\begin{figure}
\centerline{{\epsfxsize=9cm \epsfbox{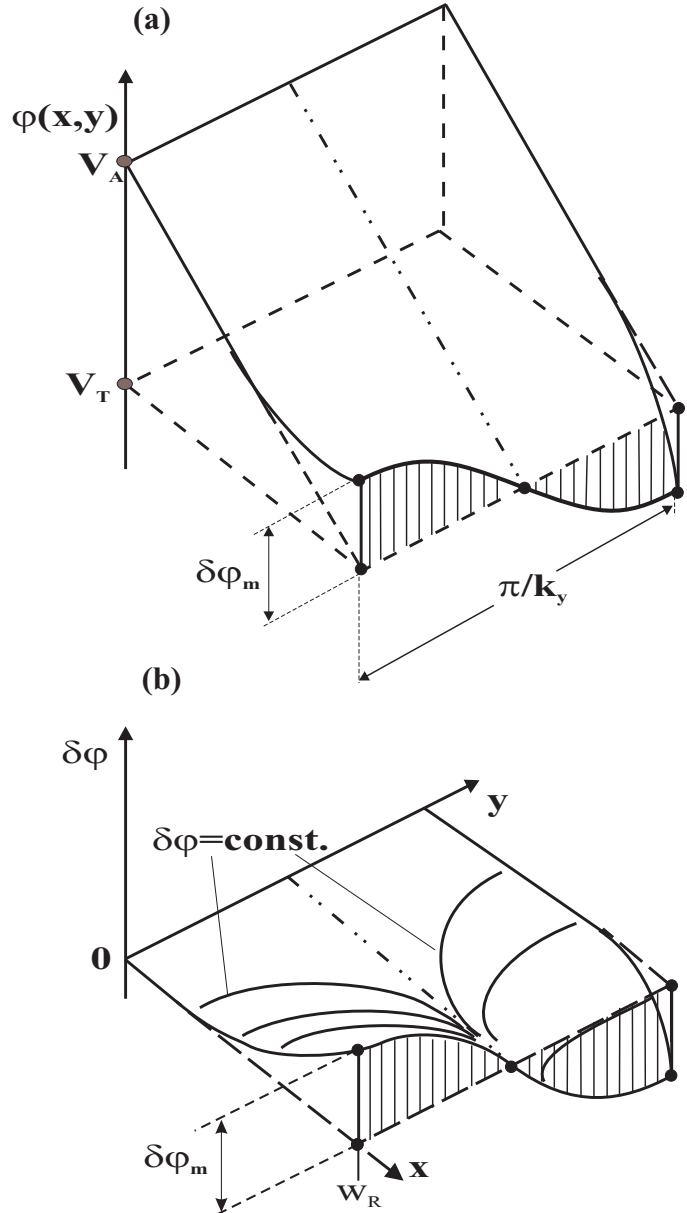}}}
\caption{Projection of the potential distribution $\varphi (x,y,z)$ 
	on the $x$-$y$ plane (a) and its
	small fluctuation $\delta \varphi$ (b) 
	in the resistive layer (schematically).}
\label{fig4}
\end{figure}

The distribution of the potential in the resistive layer,
which is essentially three-dimensional,
can be described as solution of the
Laplace equation (\ref{eq:Lapl-dgl}) with the boundary conditions
(\ref{eq:condphi}) and (\ref{eq:condphi_2}).
A general solution of this equation is a superposition
of a uniform excitation mode and various 
$k$-modes describing spatially periodic potential perturbations:
\begin{eqnarray}                                  %phiL
\label{eq:phiL}
  \varphi(x,y,z,t)=V_A - \frac{V_A-V_T}{w_R}x +
                   \sum_k B_k(t) 
                   \sinh{(k x)} \cos{(k_y y)} \cos{(k_z z)},
\end{eqnarray}
with  $k_y=m\pi /L_y, k_z=n\pi /L_z$, $k^2=k_y^2+k_z^2$, and
$m,n$ denoting integers.
Figure \ref{fig4} illustrates the potential fluctuations 
$\delta \varphi (x,y,z)$ as well as the potential 
$\varphi(x,y,z)$ in a reduced two-dimensional
space. For a given potential fluctuation
$\delta \varphi(x=w_R,y,z,t)
 = \delta\varphi_m \cos{(k_y y)} \cos{(k_z z)} \exp {(\zeta t)}$
at the boundary $w_R$ between the resistive layer and the 
thyristor-like subsystem the $k$-dependence of the potential
variations $\delta \varphi (x,y,z,t)$ in the resistive layer
is given by:
\begin{eqnarray}                                   %varphi
\label{eq:varphi}
  \delta \varphi(x,y,z,t) = \delta \varphi_m
        \frac{\sinh{(kx)}}{\sinh{(k w_R)}}
        	\cos {(k_yy)} \cos {(k_z z)} \exp {(\zeta t)}.
\end{eqnarray}

The $x$ component $j_{R,x}$ of the 
current density $\mbox{\boldmath ${j_R}$}(x,y,z,t)$
at the border between the thyristor-like subsystem 
and the resistive layer
is equal to the local anode current-density $j_T$ of
the thyristor-like subsystem [see Eqs.~(\ref{eq:Ohm}), (\ref{eq:j_T})],
so that the following relation is valid:
\begin{eqnarray}
\label{eq:JT}                                        % JT
  j_{R,x}=j_T = -\sigma_R
                \frac{\partial \varphi}{\partial x}\Bigg\vert_{x=w_R}
              = -q \mu_p N_A^-
                \frac{\partial \varphi}{\partial x}\Bigg\vert_{x=w_R}.
\end{eqnarray}
From Eqs.~(\ref{eq:varphi}) and (\ref{eq:JT}) one obtains the 
following variational derivative by which fluctuctions of
$\delta j_T$ and $\delta V_T$ are coupled for each wave-number $k$:
\begin{eqnarray}
\label{eq:deriv_R}                            % deriv_R
     \frac{\delta V_T}{\delta j_T}\Bigg\vert_k    &=&   - r_R^k,\\
   		{\mathrm with} \hspace{2em}
     r_R^k &=& \frac{w_R}{q \mu_p N_A^-}
               \frac{\tanh{(k w_R)}}{k w_R} =
               r_R^0 \frac{\tanh{(k w_R)}}{k w_R}. \nonumber
\end{eqnarray}
The quantity $r_R^k$ can be considered as an effective specific
resistance of the bulk material with respect to a 
potential fluctuation with wave-number $k$. It differs from the 
specific resistance $r_R^0 = w_R/(q \mu_p N_A^-)$ of the $p^-$ bulk
because of three-dimensional deformations of the potential relief
which arise due to a non-uniform current flow in three dimensions.

\subsection{Fluctuations in the thyristor-like subsystem}
Now we apply the linear stability analysis to the charge
dynamic equation making a perturbation Ansatz for all the variables
${\bf Y}(y,z,t)$ subjected to two-dimensional 
transversal fluctuations,
\begin{eqnarray}
\label{eq:lin}                           % lin
  {\bf Y}(y,z,t)={\bf Y}_{stat} +
         \delta{\bf Y}_m \cos(k_y y) \cos(k_z z) \exp(\zeta t),
\end{eqnarray}
and then study the stability for the both cases mentioned above.

\subsubsection{Low current range}
For the low current range the following set of variational
relations can be derived from
Eqs.~(\ref{eq:Q'-dgl})-(\ref{eq:phi_2}) and (\ref{eq:deriv_R}):
\begin{eqnarray}
\label{eq:derivs'}                           %derivs'
  \delta j_{2n} &=&
  	 \left(\zeta+\frac{1}{\tau_*}+k^2 D_n \right)~\delta Q,\\
  \delta j_{2n} &=& \left[a_C^k+
         \left(\frac{a_C^k}{r_{\cal P}^k} -
         \frac{1-a_C^k}{r_2^k}
         \right)~r_R^k \right]\delta j_T +
         \left[a_C^k k^2 D_n - (1- a_C^k)\frac{kT}{q}
         \frac{1}{r_2^k}\frac{1}{Q_{stat}}\right] \delta Q,   \\
  \delta Q 	&=&
    	-\frac{\partial Q}{\partial V_T}
                \Bigg \vert_{stat} r_R^k\delta j_T +
        \frac{\partial Q}{\partial j_{Cp}} \Bigg \vert_{stat}
                \delta j_{Cp},   \\
\label{eq:derivs'_4}                           %derivs'_4
  \delta j_{Cp} &=& (1-a_C^k)\left[1 +
        \left(\frac{1}{r_{\cal P}^k} +
        \frac{1}{r_2^k}
        \right)r_R^k\right]\delta j_T +
        (1-a_C^k) \left[k^2 D_n + \frac{kT}{q}\frac{1}{r_2^k}
        \frac{1}{Q_{stat}}\right] \delta Q,
\end{eqnarray}
where
\begin{eqnarray}
\label{eq:derQ}                             %derQ
   \frac{\partial Q}{\partial V_T}\Bigg\vert_{stat} 	=
        \tau_* \frac{a_C^0 - \alpha_1}{1-\alpha_1}\frac{1}{r_T^0}, %nonumber
   \hspace{1em}
   \frac{\partial Q}{\partial j_{Cp}}\Bigg\vert_{stat}  =
        \tau_*\frac{\alpha_1(\varphi_1)}
        {1-\alpha_1(\varphi_1)}, 	%\nonumber		  
   \hspace{1em}
   Q_{stat} = \tau_* \alpha_1 j_T,	%\nonumber
\end{eqnarray}
and the differential resistance of the thyristor-like subsystem
for the homogeneous state is given by:
\begin{equation}
        r_T^0(\varphi_1) = \frac{2V_{pth}}{j_T}\frac{\tau_*}{\tau'}
                \frac{(\alpha_1-a_C^0)}{(1-\alpha_1)^2}
                \left(
                1 - \frac{\tau_*}{\tau'}\frac{\alpha_1}{1-\alpha_1}
                \right). %
\end{equation}

The coefficient $a_C^k$ is defined as ratio of the 
fluctuations of the electron current-density $\delta j_{Cn}$ 
and the total current density $\delta j_C$ 
at the collector, which in turn can be written as:
\begin{eqnarray}
\label{eq:aka'}                                    %aka'
   a_C^k &=& \frac{\delta j_{Cn}}{\delta j_C} 
          =  \frac{dj_{1n}/d\varphi_1}{dj_1/d\varphi_1}
             \left(\frac{1}{1+r_1^{em}/r_1^k}\right) 
          =  \frac{r_1^{em}/r_{1n}^{em}}{1+r_1^{em}/r_1^k}
             \hspace{1em} {\mathrm for} \hspace{1em} 
              k \neq 0, {\mathrm and} \\
   a_C^0 &=& \frac{\delta j_{1n}}{\delta j_1}
          =  \frac{dj_{1n}/d\varphi_1}{dj_1/d\varphi_1}
          =  \frac{r_1^{em}}{r_{1n}^{em}}. \nonumber
\end{eqnarray}
The differential resistances $r_1^{em}$ and $r_{1n}^{em}$
of the $p^+$-$n^{++}$ emitter junction are defined as
\begin{eqnarray}
\label{eq:r_ks'}                                   % r_ks'
  (r_1^{em})^{-1}      &=&  \frac {d j_1}{d \varphi_1},     \\  %
  (r_{1n}^{em})^{-1}   &=&  \frac{d j_{1n}}{d \varphi_1}. 
\end{eqnarray}
The $k$-dependent quantities $r_1^k$, $r_2^k$, and $r_{\cal P}^k$
can be considered as effective resistances of the $p^+$ base, the
$n$ base, and the ${\cal P}$ layer, respectively, for a 
current-density fluctuation with wave-number $k$.

\begin{eqnarray}
\label{eq:r_ks'_2}                                   % r_ks'_2
  \left(r_1^k\right)^{-1} 
  		     &=&
            		 q \mu_p N_A^+ w_1 k^2,          \\  %
  \left(r_{\cal P}^k\right)^{-1} 
  		     &=& q \mu_p N_A^- w_{\cal P} k^2,   \\  %
  \left(r_2^k(j_T)\right)^{-1} 
                     &=& q \mu_n N_D w_2(V_T)\Bigg\vert_{stat} k^2.
% 		q b \mu_p w_{20} N_{d2}~ \frac{j_T}{j_m} k^2. %
\end{eqnarray}

Using these dependencies we obtain from 
Eqs.~(\ref{eq:derivs'})-(\ref{eq:derivs'_4})
the final dispersion relation for the increment $\zeta$
as a sum of four components:

\begin{eqnarray}
\label{eq:zet'}                                           %zet'
  \zeta (k,j_T) =
        \zeta_0 + \zeta_1  + \zeta_2 + \zeta_{\cal P}, %
\end{eqnarray}

where the partial components are:

\begin{eqnarray}
\label{eq:zets'}                                           %zets'
   \zeta_0  &=&
        	\frac{a_C^0-\alpha_1}
        	{\tau_* {\cal K}(k)}
        	\left(1+\frac{r_R^k}{r_T^0}\right) , \\ %
   \zeta_1  &=& - \frac{a_C^k r_1^{em}}{\tau_*{\cal K}(k)}
         	\times \frac{1}{r_1^k}, \\%
   \zeta_2  &=& \zeta_2'+\zeta_2'', \\ %
   \zeta_2' &=& - \frac{(1-a_C^k) r_R^k}{\tau_*{\cal K}(k)}
   		\times\frac{1}{r_2^k},               \nonumber \\%
   \zeta_2''&=&
        	- \frac{1-a_C^k}{\tau_*{\cal K}(k) \alpha_1}
        	\left[
        	\alpha_1 \left(1 + \frac {r_R^k}{r_{\cal P}^k}\right) -
        	(a_C^0-\alpha_1)\frac{r_R^k}{r_T^0}
        	\right]
        	\times \frac{kT}{q j_T}
        	\times \frac{1}{r_2^k},              \nonumber \\%
   \zeta_{\cal P}  &=&
   		 \zeta_{\cal P}' + \zeta_{\cal P}'', \\%
   \zeta_{\cal P}' &=&
   		- k^2 D_n \frac{(1-a_C^k)}{{\cal K} (k)}
        	\left[
        	\alpha_1 \left(1 + \frac {r_R^k}{r_{\cal P}^k}\right) -
        	(a_C^0-\alpha_1)\frac{r_R^k}{r_T^0}
        	\right],                              \nonumber \\%
   \zeta_{\cal P}''&=&
        	\frac{(a_C^k - \alpha_1) r_R^k}
        	{\tau_* {\cal K}(k)} \times
        	\frac{1}{r_{\cal P}^k}.               \nonumber%
\end{eqnarray}
The function ${\cal K}(k)$ is given by the expression:
\begin{eqnarray}
\label{eq:Kay}                                              %Kay
        {\cal K}(k) =
        \alpha_1 (1-a_C^k)
        \left[1 +
        \left(\frac {1}{r_{\cal P}^k}+\frac {1}{r_2^k}
        \right)~r_R^k
        \right] -
        (a_C^0-\alpha_1)\frac{r_R^k}{r_T^0}. %
\end{eqnarray}
Note that all partial components depend on the current-density dependent
coeffficients $a_c^0$, $a_c^k$ and $\alpha_1$. Besides, the contribution
of the terms $\zeta_1$, $\zeta_2$, and $\zeta_{\cal P}$ is controlled
by the effective resistances of the $p^+$ base, the $n$ base, and
the ${\cal P}$ layer, respectively. For $\zeta_{\cal P}$ also the diffusion
properties of the plasma charge play an important role.

%\newpage
\subsubsection{Moderate current range}
For the moderate current range a corresponding
set of variational relations can be obtained
by applying the linearization procedure to
Eqs.~(\ref{eq:tet1})-(\ref{eq:div2}) with taking into account
the relations Eqs.~(\ref{eq:phi_2}) and (\ref{eq:deriv_R}):
\begin{eqnarray}
\label{eq:derivs}                                    %derivs
  \delta j_{2n}	&=&
  		\left(\zeta+\frac{1}{\tau_*}+k^2 b D_p\right)~\delta Q 
        	+ \frac{b}{b+1}\left(1 + 
        	  \frac{r_R^k}{r_{\cal P}^k}\right)\delta j_T, \\%
  \delta j_{2n} &=& \left[a_C^k
        	+ \left(\frac{a_C^k}{r_{\cal P}^k} -
                \frac{1-a_C^k}{r_{2}^k}
        	\right)~r_R^k\right]\delta j_T +
        	\left[
     		 k^2 a_C^k(b-1)D_p -
        	(1-a_C^k)\frac{kT}{qQ_{stat}}\frac{1}{r_2^k}
        	\right]\delta Q,   \\%
  \delta Q 	&=& 
  		- \frac{\partial Q}{\partial V_T} \Bigg\vert_{stat}  r_R^k \delta j_T +
            	\frac{\partial Q}{\partial j_{Cp}}\Bigg\vert_{stat}  \delta j_{Cp}, \\
\label{eq:derivs_4}                                    %derivs_4
  \delta j_{Cp} &=& (1-a_C^k)\left[1+
        	\left(\frac{1}{r_2^k}+\frac{1}{r_{\cal P}^k}
        	\right)r_R^k\right]\delta j_T +
        	\left[
        	(1-a_C^k)(b-1)k^2D_p +
        	(1-a_C^k)\frac{kT}{qQ_{stat}}\frac{1}{r_2^k}
        	\right]\delta Q,%
\end{eqnarray}
where
\begin{eqnarray}
\label{eq:aka}                                          %aka
   a_C^k	
   	&=&	\frac {\alpha_1}{1+r_1^{em}/r_1^k},\\
   \frac{\partial Q}{\partial V_T}\Bigg\vert_{stat}	
        &=&	\frac{g_0\tau_*}{2r_T^0},\\ 
   \frac{\partial Q}{\partial j_{Cp}}\Bigg\vert_{stat} 
        &=&	\frac{g_0 \tau_*}{2(1-\alpha_1)}.%
\end{eqnarray}
The parameter $r_{\cal P}^k$ can again be considered as
an effective resistance of the plasma layer
with respect to current-density and potential fluctuations
with a wave-number $k$. However, in the case of
moderate currents $r_{\cal P}^k$ is conductivity modulated:
\begin{eqnarray}
\label{eq:r_pl}                                         % r_pl
    r_{\cal P}^k(Q_{stat}) 
    		  &=& \left((b+1)\mu_p Q_{stat}(j_T) k^2 \right)^{-1},\\
    Q_{stat}(j_T) &=& \tau_* g_0 j_T.
\end{eqnarray}

The final dispersion relation for the increment $\zeta$ 
results from Eqs.~(\ref{eq:derivs})-(\ref{eq:derivs_4}):
\begin{eqnarray}
\label{eq:zet}                                      %zet
   \zeta (k,j_T) = \zeta_0 + \zeta_1 + \zeta_2 + \zeta_{\cal P},
\end{eqnarray}
with
\begin{eqnarray}
\label{zets}                                         %zets
    \zeta_0 &=&	\frac{(1-\alpha_1)}{\tau_* {\cal K}(k)}
              	\left(1 + \frac{r_R^k}{r_T^0}\right),            \\%
    \zeta_1 &=& - \left(\frac{2(1-\alpha_1)}{g_0} + 1 \right)
                \frac{a_C^k r_1^{em}}{\tau_* {\cal K}(k)}
                \times \frac{1}{r_1^k},                 \\%
    \zeta_2 &=& \zeta_2' + \zeta_2'',                     \\%
    \zeta_2'&=& - \left(\frac{2(1-\alpha_1)}{g_0} + 1 \right)
              	\frac{(1-a_C^k)r_R^k}{\tau_* {\cal K}(k)}
              	\times \frac{1}{r_2^k},       \nonumber   \\%
    \zeta_2''&=& - \left[g_k
                \left( 1+\frac{r_R^k}{r_{\cal P}^k}\right) -
                (1-a_C^k)\frac{r_R^k}{r_2^k} + {\cal K}(k)\right]
                \frac{1-a_C^k}{\tau_* g_0 {\cal K}(k)}
                \frac{kT}{qj_T} \times \frac{1}{r_2^k},
                                            \nonumber   \\%
    \zeta_{\cal P} &=& \zeta_{\cal P}' +  \zeta_{\cal P}'', \\%
    \zeta_{\cal P}'&=& -k^2 D_p
                     \left\{b(1-a_C^k) + a_C^k +
                     \frac{(b-1)(1-a_C^k)}{{\cal K}(k)}
                     \left[ g_k\left(1+ \frac{r_R^k}{r_{\cal P}^k}
                     \right)-
                     (1 - a_C^k)\frac{r_R^k}{r_2^k}
                     \right] \right\},       \nonumber    \\%
    \zeta_{\cal P}''&=& \left[2 g_k(1-\alpha_1) -
                      g_0(1-a_C^k) \right]
                      \frac{r_R^k}{\tau_* g_0 {\cal K}(k)}
                      \times \frac{1}{r_{\cal P}^k},
                                             \nonumber
\end{eqnarray}
and
\begin{eqnarray}
	g_k = a_C^k - \frac{b}{b+1},
\end{eqnarray}
\begin{equation}                                %Kay_M
\label{Kay_M}
    {\cal K}(k) = (1-a_C^k)
                  \left(1 + \frac{r_R^k}{r_2^k} +
                  \frac{r_R^k}{r_{\cal P}^k}  \right) -
                  (1 - \alpha_1) \frac{r_R^k}{r_T^0} > 0.
\end{equation}
Thereby the differential resistance $r_T^0$ of the thyristor-like
subsystem is determined by Eq.~(\ref{eq:r_T}).

For both the low and the moderate current case
the increment $\zeta (k)$ is proportional to the sum
of the differential resistance of the thyristor-like
subsystem, $r_T^0$, and that of the resistive layer, $r_R^0$,  
in the limit of perturbations with small wave-numbers:
\begin{eqnarray}
\label{eq:limit}                                          %limit
  \zeta (k \to 0 ) \propto -(r_R^0+r_T^0). %
\end{eqnarray}
This obviously corresponds to a one-dimensional loading of the
thyristor-like subsystem  by the resistive layer.
From Eq.~(\ref{eq:limit}) follows the well-known criterium
for stability with respect to homogeneous fluctuations:
\begin{eqnarray}
\label{eq:stabcrit}                                          %stabcrit
	r_R^0 + r_T^0 > 0.
\end{eqnarray}

\subsection{Dispersion relations and discussion}

\begin{figure}
\centerline{{\epsfxsize=9cm \epsfbox{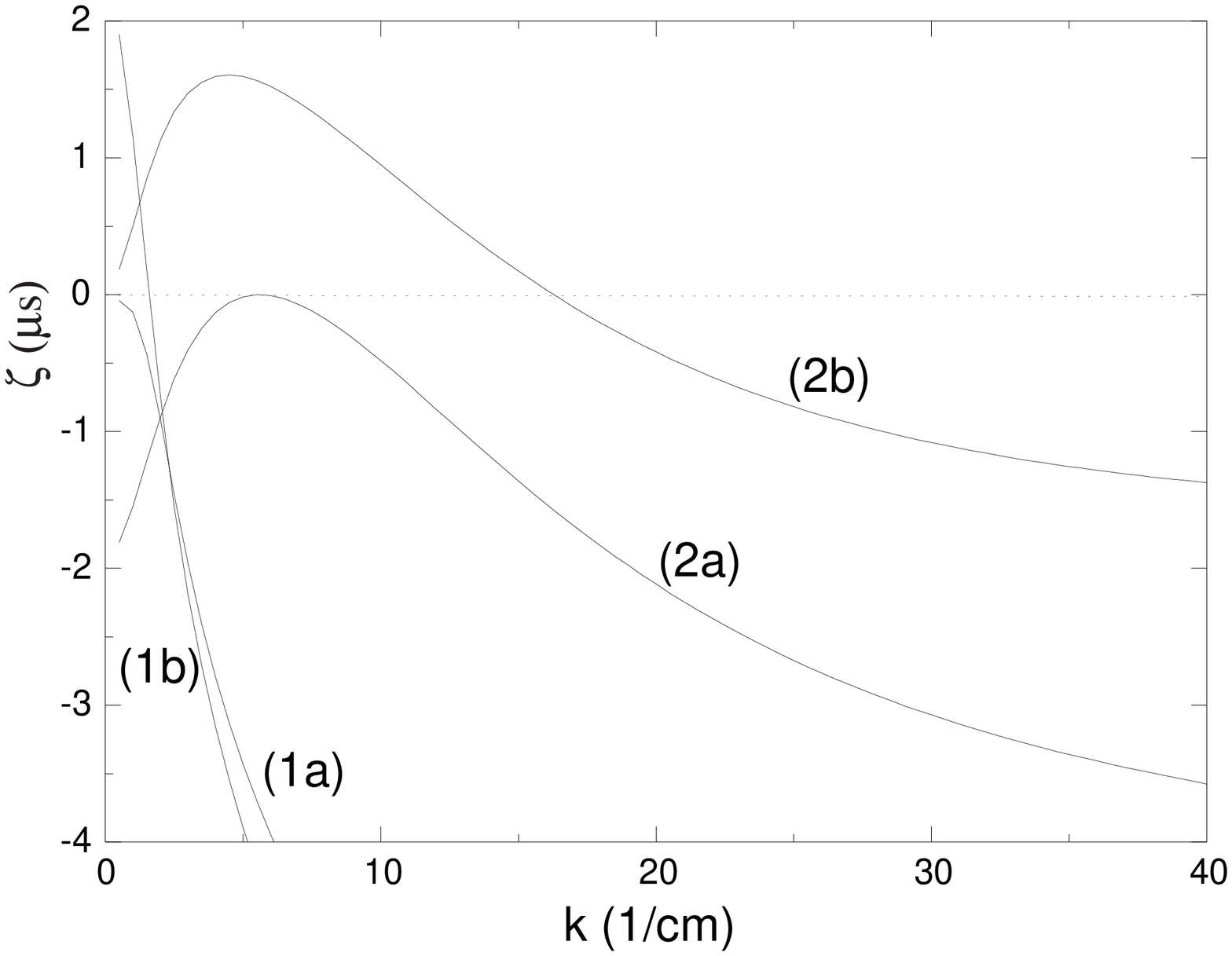}}}
\caption{Dispersion relations $\zeta (k)$ for the low current case
	revealing the instability of the uniform state with respect
	to large wave-length fluctuations, $k \to 0$, (1) and with
	respect to
	a fluctuation with a critical wave-number $k > 0$ (2);
	$\sigma_1^{eff} = 0.04 \sigma_1, 5 \cdot 10^{-4} \sigma_1$, 
	for curves (1) and (2), and 
	$j = 1.01, 0.671, 1.11, 1.01$ mA/cm$^2$ 
	for curves 1a, 1b, 2a, 2b, respectively.
	Other parameters as in Table \protect{\ref{table2}}.}
\label{fig5}
\end{figure}

Figure \ref{fig5} shows 
dispersion relations $\zeta(k)$ for the low current case. 
The calculations have been performed with two reduced values
of the $p^+$ base conductivity $\sigma_1^{eff}$ in order to 
illustrate two qualitatively different instabilities that
can occur in the semiconductor structure. 
If the $p^+$ base conductivity is sufficiently low, dispersion
relations like the curves denoted by (2) can occur. They are
characterized by a pronounced maximum at a critical 
wave-number that is definitely larger than zero. 
Curve (2a) represents the case that the increment $\zeta$ of this
critical wave-number is zero, while all other increments are still
smaller than zero, i.\,e., the uniform state becomes
destabilized with respect to current-density fluctuations with
the critical wave-number. When the sample current is decreased,
the curve is shifted up [curve (2b)], and the increments of a 
certain band of wave-numbers are larger than zero. 
This behavior can be interpreted as a 
Turing-like instability \cite{Tur}.

When the transversal coupling inside the thyristor-like subsystem
is strengthened, 
e.\,g. by reducing the transversal resistance of the $p^+$ base,
the maxima of $\zeta(k)$ shift to smaller $k$ values
leading to dispersion relations like those denoted by (1).
They are characterized by a monotonous decrease of $\zeta(k)$ 
with growing $k$, which can be considered as an 
instability of Ridley's type \cite{Rid}. In such a case, 
the uniform state is typically destabilized via
a saddle node bifurcation by 
fluctuations with the shortest possible wave-number 
$2\pi/(L_y^2 + L_z^2)$, if uniform fluctuations are suppressed
by a sufficiently large load resistor in the external
circuit.
Similar to the Turing-like case, the curve is shifted upwards,
when the sample current is decreased. Note that the parameter
$j$ is the same for the curves (1a) and (2b) and has been adjusted such
that the differential resistance of the global current-density
vs.~voltage characteristic vanishes, i.\,e., $r_T^0 + r_R^0 = 0$.

\begin{figure}
\centerline{{\epsfxsize=9cm \epsfbox{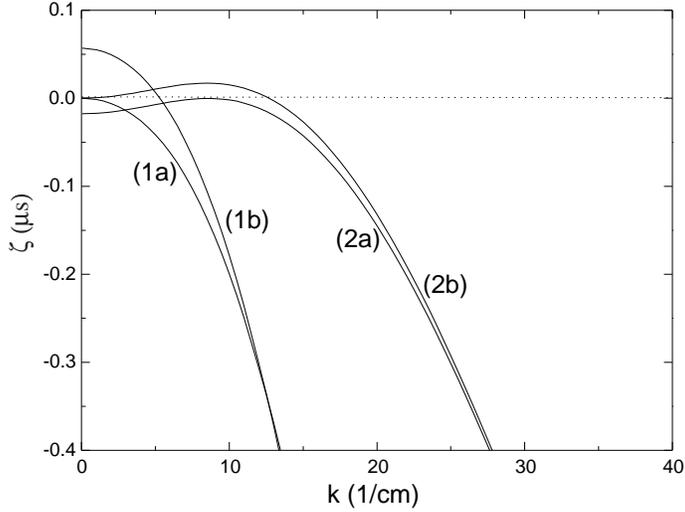}}}
\caption{Dispersion relations $\zeta (k)$ for the moderate current case
	revealing the instability of the uniform state with respect
	to large wave-length fluctuations, $k \to 0$, (1) and with
	respect to
	a fluctuation with a critical wave-number $k > 0$ (2);
	$\sigma_1^{eff} = 0.25 \sigma_1, 
	0.05 \cdot 10^{-4} \sigma_1$, 
	for curves (1) and (2), and 
	$j = 394, 350, 407, 394 $ mA/cm$^2$ for curves 1a, 1b, 2a, 2b,
	respectively. Other parameters as in Table 
	\protect{\ref{table2}} besides 
	$N_A^- = 5 \cdot 10^{13}$ cm$^{-3}$ and $w_R = 0.1$ cm.}
\label{fig6}
\end{figure}

Dispersion relations $\zeta(k)$ for the moderate current case are
shown in Fig.~\ref{fig6}. Similar to the low current case the
homogeneous state is destabilized at a critical current level either
by an instability of Ridley's (curves 1) or Turing's type (curves 2). 
The bifurcation type can be controlled by a proper design. 
The most important parameters are the
conductivities of the resistive layer, the $p^+$ and the $n$ base. 
The former essentially governs the transversal spreading of the 
potential drop across the resistive layer and tends to damp
out current-density fluctuations. The $p^+$ and
$n$ base conductivity control the transversal hole and 
electron current densities
in the two base layers. As these base currents in 
turn regulate the injection of electrons and holes from the
$n^+$ emitter and the plasma layer $\cal P$, respectively, they
play an important role concerning the transversal spreading
of a region with enhanced current density. 
The competition between the damping and the activating process,
is the essential physical mechanism allowing 
- under properly chosen design parameters - the
destabilization of a homogeneous current-density distribution 
by spatially periodic current-density fluctuations strongly
reminding of the generic mechanism proposed by Turing.

\section{Summary}
In conclusion we point out that the suggested theory,
which is based on a subdivision of the semiconductor structure
into an active and a passive subsystem, supplies a self-consistent
quantitative description of the nonlinear mechanisms, which control
the longitudinal current flow, and the transversal
current-density instabilities in multilayer semiconductor systems.
The nonlinear longitudinal electrical properties
in the active layer are based on a thyristor-like regenerative
mechanism, while the passive subsystem acts as a simple resistive
layer. Depending on the system parameters two different 
types of regenerative mechanisms can be distinguished leading
to specific model equations. 
A stability analysis of the derived equations reveals that in each
case two basic bifurcation types can be expected when the
total sample current is varied. The bifurcations are caused by
instability mechanisms that are closely related to those studied
by Turing \cite{Tur} and Ridley \cite{Rid}. It turned out that
in particular the conductivity ratios of the $p^+$ base, the 
plasma layer, and the resistive layer are important parameters
to control the instability mechanism. Thus, adjusting the
respective transversal conductivities, e.\,g., by irradiating
the sample with protons or ions, using compensated substrate material,
or adapting the sample geometry, it is possible to design
a multilayer system with the aspired (in)stability features.
Finally, we emphasize that the suggested approach presents a
basis for further numerical analysis of pattern formation
processes.

\section*{Acknowledgements}
The present paper was prepared in the frame of the international
project "Dissipative Pattern Formations in Semiconductors
and Semiconductor Devices" supported by the Deutsche
Forschungsgemeinschaft and the Russian Academy of Science.

\renewcommand{\baselinestretch}{0.95}
%

%***********************************       T A B L E   1     *********************
\newpage
\begin{table}
\caption{Physical constants  and material parameters}
\vspace{2pt}
\begin{tabular}{cp{10cm}c}
%\hline
\multicolumn{1}{c}{\rule{0pt}{12pt} Symbol}
	&\multicolumn{1}{c}{Symbol meaning}
	&\multicolumn{1}{c}{Value}\\[2pt]
\hline\rule{0pt}{12pt}
%Parameter	& Parameter meaning		& value \\[0.5ex]
%
$b$		& $b=D_n/D_p$			& 2.8 \\
$D_n$		& diffusion coefficient of 
		  electrons 			& $D_n = kT \mu_n /q$ \\	
$D_p$		& diffusion coefficient of 
		  holes 			& $D_p = kT \mu_p /q$ \\	
$k$		& Boltzmann constant	& $8.62 \cdot 10^{-5}$V/K\\
$n_i$		& intrinsic carrier concentration & $1.45 \cdot 10^{10}$ cm$^{-3}$ \\
$q$		& elementary charge		
					& $1.6 \cdot10^{-19}$ As\\ 
$\epsilon_0$	& vaccuum permittivity		
					& $8.854 \cdot 10^{-14}$  F/cm\\
$\epsilon$	& dielectric constant of Si	& 11.7\\
$\mu_n$		& electron mobility		& $b \mu_p$ \\
$\mu_p$		& hole mobility			& 480 cm$^2$/Vs\\
%\hline
\end{tabular}
\label{table1}
\end{table}

%***********************************       T A B L E   2     *********************
%\newpage
\begin{table}
\caption{Set of design parameters}
\vspace{2pt}
\begin{tabular}{cp{10cm}c}
%\hline
\multicolumn{1}{c}{\rule{0pt}{12pt} Symbol}
	&\multicolumn{1}{c}{Symbol meaning}
	&\multicolumn{1}{c}{Value}\\[2pt]
\hline\rule{0pt}{12pt}
%Parameter	& Parameter meaning		& Standard parameter value \\[0.5ex]
$j_{1ns}$	& saturation value of the
		  electron injection current-density
		  at the $p^+$-$n^{++}$ junction     
		  			& $2 \cdot 10^{-11}$ Acm$^{-2}$ \\
$j_{1ps}$	& saturation value of the
		  hole injection current-density
		  at the $p^+$-$n^{++}$ junction	
		  			& $1 \cdot 10^{-13}$ Acm$^{-2}$ \\
$j_{1Rs}$	& saturation value of the 
		  recombination-generation current-density
		  at the $p^+$-$n^{++}$ junction	
		  			& $1 \cdot 10^{-8}$ Acm$^{-2}$ \\
$N_A$		& acceptor concentration 
		  of the $p^+$ emitter	& $1 \cdot 10^{18}$ cm$^{-3}$\\
$N_A^-$		& acceptor concentration 
		  of the $p^-$ bulk		
		  			& $1 \cdot 10^{13}$ cm$^{-3}$\\
$N_A^+$		& acceptor concentration 
		  of the $p^+$ base		
		  			& $5 \cdot 10^{17}$ cm$^{-3}$\\
$N_D$		& donor concentration 
		  of the $n$ base		
		  			& $1.2 \cdot 10^{14}$ cm$^{-3}$\\
$N_D^+$		& donor concentration 
		  of the $n^{++}$ emitter	
		  			& $1 \cdot 10^{20}$ cm$^{-3}$\\
$w_1$		& width of the $p^+$ base	& $2 \cdot 10^{-4}$ cm\\
$w_{20}$	& width of the $n$ base	
		  for $V_C = 0$			& $1.5 \cdot 10^{-3}$ cm\\
$w_{\cal{P}}$	& width of the plasma layer	& $5 \cdot 10^{-3}$ cm\\
$w_R$		& width of the resistive layer	& 1 cm\\
$\sigma_1$	& conductivity of the $p^+$ base
			& $38.4 (\Omega cm)^{-1}$\\
$\sigma_2$	& conductivity of the $n$ base  
			& $2.58 \cdot 10^{-2} (\Omega cm)^{-1}$\\
$\sigma_{\cal{P}}$
		& conductivity of the 
		  $\cal{P}$ layer  		  	& \\
$\sigma_R$	& conductivity of the $p^-$ bulk  
			& $7.68 \cdot 10^{-4} (\Omega cm)^{-1}$\\
$\tau$		& lifetime in the $\cal{P}$ layer
		  for low currents			& $1 \mu$s\\
$\tau_h$	& lifetime in the $\cal{P}$ layer
		  for moderate currents			& $5 \mu$s\\
%\hline
\end{tabular}
\label{table2}
\end{table}

%%%%%%%%%%%%%%%%%%%%%%%%%%%%%    F I G U R E S      %%%%%%%%%%%%%%%%%%%%%%%%%%%%%
\newpage

\newpage

\end{document}